# Efficient and reliable method for the simulation of scanning tunneling images and spectra with local basis sets

Óscar Paz* and José M. Soler†
*Dpto. Física de la Materia Condensada, C-III, Universidad Autónoma de Madrid, E-28049 Madrid, Spain*

Based on Bardeen's perturbative approach to tunneling, we have found an expression for the current between tip and sample, which can be efficiently coded in order to perform fast *ab initio* simulations of STM images. Under the observation that the potential between the electrodes should be nearly flat at typical tunnel gaps, we have addressed the difficulty in the computation of the tunneling matrix elements by considering a vacuum region of constant potential delimited by two surfaces (each of them close to tip and sample respectively), then propagating tip and sample wave functions by means of the vacuum Green's function, to finally obtain a closed form in terms of convolutions. The current is then computed for every tip-sample relative position and for every bias voltage in one shot. The electronic structure of tip and sample is calculated at the same footing, within density functional theory, and independently. This allows us to carry out multiple simulations for a given surface with a database of different tips. We have applied this method to the Si(111)-(7×7) and Ge(111)-*c*(2×8) surfaces. Topographies and spectroscopic data, showing a very good agreement with experiments, are presented.

## I. INTRODUCTION

Scanning tunneling microscopy [1] (STM) has opened the door to a deep knowledge about metal and semiconducting surfaces; thus, revolutionizing the way in which their features are investigated [2, 3]. The exponential dependence of the current, as a sharp tip approaches to the sample, was technically exploited to achieve high vertical (∼ 0.1 Å) and lateral (∼ 1 Å) spatial resolutions, providing quite frequently a detailed three-dimensional representation (i.e., atomic resolution) of the apparent topography of the sample. An early success was the clear identification of the Si(111)-(7×7) structure [4, 5]. The experimental challenge was even greater in scanning tunneling spectroscopy (STS) measurements, where the energy spectrum of the sample is locally probed. Real-space images of electron states and their dependence with energy can be observed by tuning the bias voltage applied between the electrodes for a concrete relative tip-sample position. In careful measurements, the energetic resolution at very low temperatures can reach ∼ 10 mV. Therefore, it is not surprising that STM has currently became an essential research tool in surface science; very suitable in the study of both electronic and structural local properties. Applications range from the research on atomic arrangements and reconstructions of pristine surfaces or the characterization of surface defects or adsorbates on them, to the study of electronic properties (like charge-density waves), vibrational properties or dynamical processes (like diffusion or oxidation), to give only some examples.

STM images are very often understood as a contour map of the surface local density of states (LDOS) of the sample at the Fermi level. This interpretation was firstly proposed by Tersoff and Hamann (TH) [6, 7], who developed a simple formula for the tunneling current in the case of a probe with a maximum resolution, but without a realistic chemical structure. When the experiments are highly reproducible, regardless of the tip used, this can be sufficient. However, the information conveyed by the experimental data is much more rich and complex than the picture of this approximation, because they involve the convolution of both sample and tip states. An intrinsic inconvenience associated with these techniques is that the geometrical shape and composition of the tip is unknown during the operation. In fact, it is also safe to think that the tip is even rarely in equilibrium, due to uncontrolled tip-sample interactions or eventual contacts with the surface that could entirely modify its structure. These uncertainties are particularly crucial in the case of STS, where slight changes in the tip can give rise to a completely different spectra. As a final consequence, it is common that experiments have a low experimental reproducibility, and that conclusions from a direct observation of the experimental information could not be soundly extracted in all circumstances.

An ordinary strategy to gain insight in the interpretation of measurements is to perform careful comparisons with first-principles simulations in which multiple configurations of the tip are employed. To this end, many theoretical approaches have been proposed, starting from the simplest and widely used Bardeen's formalism [8], specially in its TH formulation [6, 7]. Within the latter approximation, it is possible to obtain a qualitative agreement with experiments in some cases, but it is generally not so easy to attain quantitative data or to provide explanations to many observations, such as bias-dependent images or negative differential resistances. This situation has favored, on the one hand, the development of new methods which increasingly include the probe composition [9–14]. On the other hand, even more sophisticated approaches, which treat the interaction between the electrodes more thoroughly, have used nonperturbative expressions for the tunneling current [15–22] based on the Landauer-Buttiker [23, 24] and the nonequilibrium Green's function formalism (also known as the Keldysh [25] or the Kadanoff-Baym [26] formalism). These schemes are important at close tip-sample distances (< 5 Å), where multiple scattering effects or even tip-sample contact are present, but not in the standard tunneling regime (5 – 10 Å) of most experiments. Under these conditions, the first order contributions to the current are dominant, and many relevant features in the images can be captured starting from the electronic structure of isolated electrodes. Nevertheless, a systematic and reliable analysis of the STM images has been limited by the balance between the efficiency and the accuracy. Some topographic simulations demand large amounts of computer time, specially to obtain also spectroscopic data. Therefore, simulations for a given sample and several tip compositions or structures were not feasible with a reasonable effort. In the present work, a method aimed at performing real time and reliable simulations is developed. It could be classified among the family of the perturbative approaches, therefore being applicable under the limitations commented above. Otherwise it incorporates the band structure of tip and sample at the same level of theory, using density-functional theory (DFT), and allows multiple simulations for different set-points (tunneling current and bias voltage) and for several tips, without the need of high computational resources.

*Electronic address: Oscar.Paz@UAM.es
†Electronic address: Jose.Soler@UAM.es

## II. THEORETICAL FRAMEWORK

A pioneering advance to the theory of tunneling was developed by Bardeen [8], who used time-dependent perturbation theory to describe the tunneling process by a small coupling between two independent electrodes. On the basis of the Fermi's golden rule, the tunnel current can be expressed as

$$I = \frac{2\pi e}{\hbar} \sum_{t,s} [f(\varepsilon_t) - f(\varepsilon_s)]\, |M_{ts}|^2\, \delta(\varepsilon_t - \varepsilon_s + eV). \qquad (1)$$

Here $f(\varepsilon_j)$ is the Fermi-Dirac function, the energies $\varepsilon_j$ ($j = t, s$) are referred to the Fermi levels of the tip and sample respectively, and $V$ is the applied bias voltage between the electrodes. The Bardeen matrix element, $M_{ts}$, which couples state $\varphi_t$ (with energy $\varepsilon_t$) of the isolated tip and state $\varphi_s$ ($\varepsilon_s$) of the isolated sample, quantifies the tunnel probability and takes the following form,

$$M_{ts} = -\frac{\hbar^2}{2m} \int_\Sigma [\varphi_t^*(\boldsymbol{r})\nabla\varphi_s(\boldsymbol{r}) - \varphi_s(\boldsymbol{r})\nabla\varphi_t^*(\boldsymbol{r})] \cdot d^2\hat{\boldsymbol{r}}, \qquad (2)$$

where the integral can be calculated over *any* surface lying in the vacuum region between the two electrodes. The total tunnel current will be therefore a sum over all states in the energy window delimited by the bias voltage, and under the condition of elastic tunneling, stated by the delta function in Eq. (1). Although this approach is in principle a sound, simple, and accurate model for tunneling, the simulation of STM images and STS spectra from first-principles calculations has been burdened by technical difficulties.

### A. Propagation of wave functions

The most difficult part in the evaluation of Eq. (1) emerges from the computation of $M_{ts}$. Even if the surface of integration could be chosen as simple as possible (for instance as an intermediate plane in the vacuum zone, and the integral computed in a regular grid of points), it is interesting to emphasize that *ab initio* wave functions and their gradients are very tricky to converge properly in that region. The first technical problem is intrinsically related with the presence of the vacuum region, where wave functions are very small and energetically irrelevant with respect to the bulk. Consequently an extremely good self-consistent convergence is required to obtain accurate values of wave functions far away from both surfaces. Moreover, a second drawback arises due to the imperfect description of wave functions in that zone, since basis sets are not spatially complete enough, which becomes specially critical in the case of atom-centered basis sets, like atomic orbitals. Since these difficulties increase with the tip-sample distance, *ab initio* simulations are commonly performed at close and unrealistic separations, thus complicating a safe comparison with measurements performed at typical tunnel gaps of 5 – 10 Å. An additional technical inconvenient is the need of the inclusion of both tip and sample in the same simulation cell, which forces to treat the system as a whole and to perform the computation of $M_{ts}(\boldsymbol{R})$ for every tip position $\boldsymbol{R}$, giving rise to a significant penalty in the efficiency of the calculations.

The TH model [6, 7] can be applied to eliminate the tip uncertainties and decouple the computation of the whole tip-sample system. In the TH approximation the potential between the tip and sample is considered to be constant, which leads to

$$\nabla^2 G_t(\boldsymbol{r}-\boldsymbol{R}) - \kappa_t^2 G_t(\boldsymbol{r}-\boldsymbol{R}) = -\delta(\boldsymbol{r}-\boldsymbol{R}), \text{ with } \kappa_t^2 = \frac{2m}{\hbar^2}(\phi_t - \varepsilon_t), \qquad (3)$$

that is, the equation describing a punctual tip located at $\boldsymbol{R}$ with a continuous energy spectrum formed by spherically symmetric states. The solution to the above expression, satisfying the contour conditions, is of the form

$$G_t(\boldsymbol{r}-\boldsymbol{R}) = G_t(\rho) = \frac{e^{-\kappa_t \rho}}{4\pi\rho}, \text{ with } \rho = |\boldsymbol{r}-\boldsymbol{R}|. \qquad (4)$$

Also, the value of the work function of the tip, $\phi_t$, has been explicitly introduced in Eq. (3). The dependence of the subsequent simulated STM images on variations in this quantity or other parameters defined hereafter will be discussed in section II F.

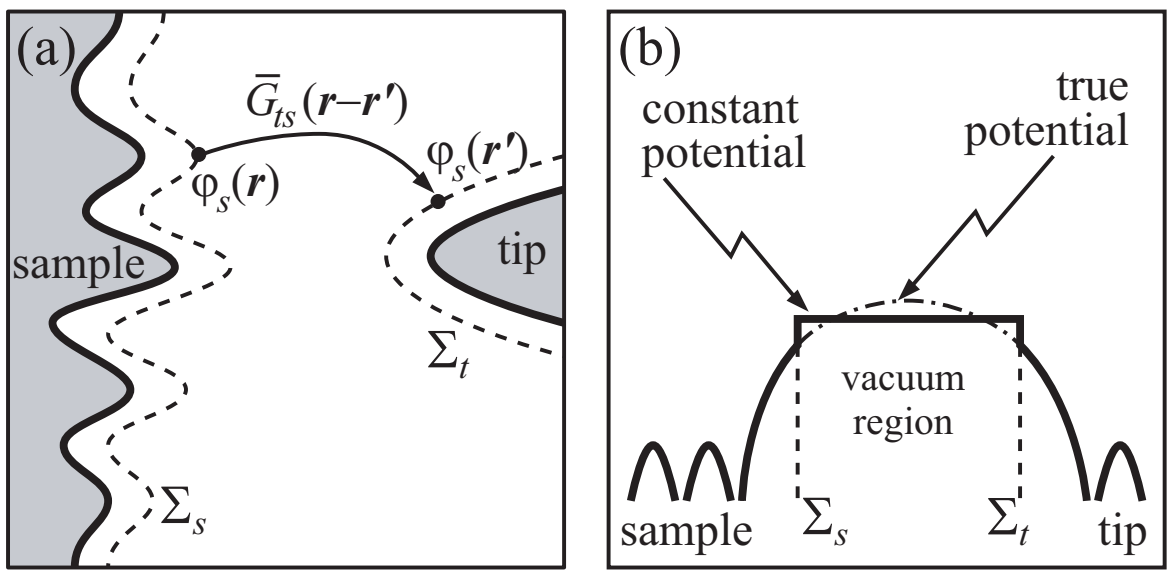

FIG. 1: (a) A schematic representation of the propagation of $\varphi_s(\boldsymbol{r})$ values at the mathematical surface $\Sigma_s$ across the vacuum region up to the points $\boldsymbol{r}'$ in $\Sigma_t$, using the modified vacuum Green's function $\bar{G}_{ts}(\boldsymbol{r}-\boldsymbol{r}')$ [see Fig. 2]. $\Sigma_s$ ($\Sigma_t$) denotes an isosurface of the electron density in the proximities of the sample (tip) [refer to Eq. (6)]. (b) The approximation of a flat effective potential in the gap between the tip-sample system.

The tunneling matrix elements of Eq. (2) can now be straightforwardly evaluated, using as the tip state the Green's function in Eq. (3) and applying the Green's theorem afterwards, which leads to

$$M_{ts}(\boldsymbol{R}) = -\frac{\hbar^2}{m}\sqrt{2\pi\kappa_t}\int_{\Sigma}[G_t^*(\boldsymbol{r}-\boldsymbol{R})\nabla\varphi_s(\boldsymbol{r})+ \\ -\varphi_s(\boldsymbol{r})\nabla G_t^*(\boldsymbol{r}-\boldsymbol{R})]\cdot d^2\hat{\boldsymbol{r}} = -\frac{\hbar^2}{m}\sqrt{2\pi\kappa_t}\,\varphi_s(\boldsymbol{R}). \tag{5}$$

And it is easy to show that the tunneling current reduces to the LDOS of the sample near the Fermi energy and located at the tip position, $I \propto \int_{\varepsilon_F}^{\varepsilon_F+eV}\rho_s(\boldsymbol{R},\varepsilon)\,d\varepsilon$. A clear advantage of such expression is the simplicity in its implementation, since the electronic structure of the tip does not contribute at all. Conversely, this interpretation generally allows only for qualitative results, since the matching between experiments and simulations is done by tuning the isovalue of the LDOS, which can be strongly sensitive in many cases. Moreover, the quantity $\rho_s(\boldsymbol{R},\varepsilon)$ also requires both good basis set and self-consistent convergence in the calculations to realistically represent its exponential decaying at large distances.

Alternatively, Eq. (5) can be used as a mathematical recipe, *in the opposite direction*, to obtain the value of $\varphi_s$ everywhere. In particular we can propagate the wave functions from the positions $\boldsymbol{r}$ of a surface $\Sigma_s$, close to the sample, up to the points $\boldsymbol{r}'$ of another surface $\Sigma_t$, in the surrounding area of the tip (as illustrated in Fig. 1). There we can then substitute these accurate wave functions in Eq. (2), to find more precise values for the matrix elements. Although this implies the calculation of two surface integrals, it clearly overcomes the technical difficulties related with the vacuum region, while providing a good numerical accuracy. Indeed, since the method is completely symmetric, it can be also seen as the propagation of the tip wave functions from $\Sigma_t$ to $\Sigma_s$. For that reason, we apply the Green's function $\bar{G}_{ts}$, which depends on both tip and sample states. Its functional form will be introduced and commented in section II C.

### B. Broadening of surface integrals

In the evaluation of $M_{ts}$, the integration can be greatly simplified if $\Sigma$ is chosen to be a plane. However, this geometry cannot faithfully adapt to the topology of the sample, especially for highly corrugated cases like surfaces presenting terraces, defects or adsorbates on them. A better choice for $\Sigma_t$ and $\Sigma_s$ is as isosurfaces of constant electron density,

$$\Sigma: \; S(\boldsymbol{r}) = \log\big(\rho(\boldsymbol{r})/\rho_0\big) \equiv 0, \tag{6}$$

where $\rho_0$ is a parameter indicating a value of reference. It should be chosen so that the isosurfaces are close enough to the physical surfaces, to ensure precise first-principles wave functions, and far enough to make adequate the constant potential approximation. The choice of such mathematical surface, although physically more convenient, complicates the computation of the surface integrals. To deal with those geometries, we transform them into volume integrals by constraining the volume element in the local perpendicular direction to the isosurfaces, $u(\boldsymbol{r}) = \boldsymbol{r}\cdot\nabla S(\boldsymbol{r})/|\nabla S(\boldsymbol{r})|$, so that

$$\int_{\Sigma}\boldsymbol{f}(\boldsymbol{r})\cdot d^2\hat{\boldsymbol{r}} = \int \boldsymbol{f}(\boldsymbol{r})\cdot\frac{\nabla\rho(\boldsymbol{r})}{|\nabla\rho(\boldsymbol{r})|}\,\delta\big(u(\boldsymbol{r})\big)\,d^3r = \int \boldsymbol{f}(\boldsymbol{r})\cdot\boldsymbol{c}(\boldsymbol{r})\,d^3r, \tag{7}$$

with $\boldsymbol{c}(\boldsymbol{r}) = \delta\big(S(\boldsymbol{r})\big)\,\nabla\rho(\boldsymbol{r})/\rho(\boldsymbol{r})$. The delta in the constraint function $\boldsymbol{c}(\boldsymbol{r})$ can be appropriately broadened to ensure that the surface is well represented by the mesh points. In particular we use the function

$$\delta\big(S(\boldsymbol{r})\big) = \begin{cases} \frac{15}{16\Delta S}\left[1-\left(\frac{S(\boldsymbol{r})}{\Delta S}\right)^2\right]^2 & \text{if } -\Delta S < S(\boldsymbol{r}) < \Delta S, \\ 0 & \text{elsewhere,} \end{cases} \tag{8}$$

where $\Delta S$ is set considering the decay of the electronic density and the grid spacing.

Once the mathematical surface is broadened, Eq. (5) can be very efficiently implemented, since it can be expressed as a three-dimensional convolution with the help of Green's theorem,

$$\varphi_s(\boldsymbol{r}') = \int [G_t^*(\boldsymbol{r}-\boldsymbol{r}')A_s(\boldsymbol{r}) - \boldsymbol{B}_s(\boldsymbol{r})\cdot\nabla G_t^*(\boldsymbol{r}-\boldsymbol{r}')]\,d^3r = \\ = \frac{1}{(2\pi)^{3/2}}\int \tilde{g}_t^*(\boldsymbol{k})\,[\tilde{A}_s(\boldsymbol{k}) + i\boldsymbol{k}\cdot\tilde{\boldsymbol{B}}_s(\boldsymbol{k})]\,e^{i\boldsymbol{k}\cdot\boldsymbol{r}'}\,d^3k, \tag{9}$$

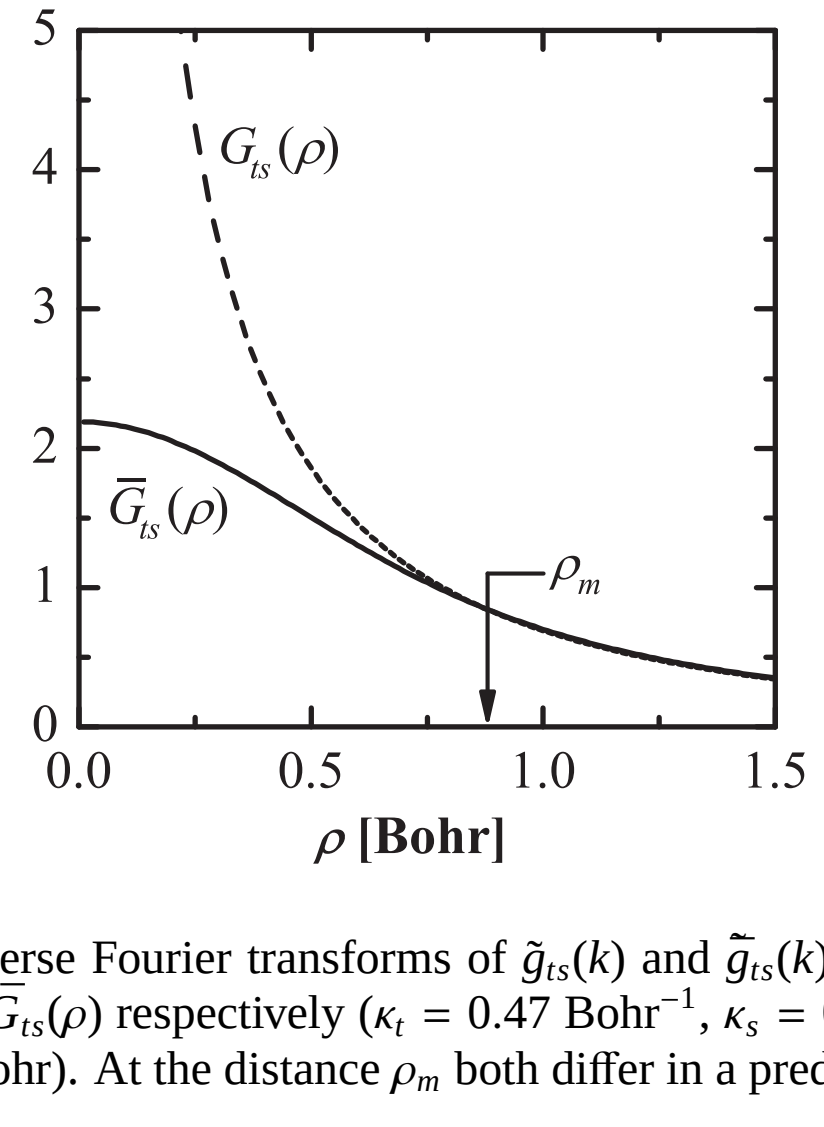

FIG. 2: Inverse Fourier transforms of $\tilde{g}_{ts}(k)$ and $\bar{\tilde{g}}_{ts}(k)$, denoted as $G_{ts}(\rho)$ and $\bar{G}_{ts}(\rho)$ respectively ($\kappa_t = 0.47$ Bohr$^{-1}$, $\kappa_s = 0.72$ Bohr$^{-1}$, $\alpha = 0.25$ Bohr). At the distance $\rho_m$ both differ in a predefined tolerance.

where we make use of the following definitions

$$\begin{aligned} A_j(\boldsymbol{r}) &= \boldsymbol{c}_j(\boldsymbol{r})\cdot\nabla\varphi_j(\boldsymbol{r}) = \frac{1}{(2\pi)^{3/2}}\int \tilde{A}_j(\boldsymbol{k})\, e^{i\boldsymbol{k}\cdot\boldsymbol{r}}\, d^3k, \\ \boldsymbol{B}_j(\boldsymbol{r}) &= \boldsymbol{c}_j(\boldsymbol{r})\,\varphi_j(\boldsymbol{r}) = \frac{1}{(2\pi)^{3/2}}\int \tilde{\boldsymbol{B}}_j(\boldsymbol{k})\, e^{i\boldsymbol{k}\cdot\boldsymbol{r}}\, d^3k, \\ G_j(\boldsymbol{r}-\boldsymbol{r}') &= \frac{1}{(2\pi)^{3/2}}\int \tilde{G}_j(\boldsymbol{k};\boldsymbol{r}')\, e^{i\boldsymbol{k}\cdot\boldsymbol{r}}\, d^3k. \end{aligned} \tag{10}$$

### C. Modified vacuum Green's function

The vacuum Green's function in reciprocal space, $\tilde{G}_t(\boldsymbol{k};\boldsymbol{r}') = \tilde{g}_t(\boldsymbol{k})\, e^{-i\boldsymbol{k}\cdot\boldsymbol{r}'} = 1/(k^2+\kappa_t^2)\, e^{-i\boldsymbol{k}\cdot\boldsymbol{r}'}$, can be simply obtained from the Fourier transform of Eq. (3). However, $\tilde{g}_t$ depends on the work function $\phi_t$ through $\kappa_t$, and the work functions of the tip and sample are generally different. Therefore a function in which $\phi_t$ and $\phi_s$ were averaged would be a better approximation, since the roles of tip and sample should be interchangeable. The simplest way is to evaluate $\tilde{g}_t$ for the mean of $\phi_t$ and $\phi_s$. Nevertheless, it is more convenient to average the $\tilde{g}_t$ and $\tilde{g}_s$ functions themselves for implementation purposes. We use the geometrical mean in reciprocal space $\tilde{g}_{ts}(k) = \sqrt{\tilde{g}_t(k)\,\tilde{g}_s(k)}$, although other schemes (e.g., arithmetic mean, or geometric mean in real space) would give similar results. We then also modify the product as: $\tilde{\bar{g}}_{ts}(k) = \tilde{g}_{ts}(k)\, e^{-\alpha^2k^2}$, where we have used a Gaussian filtering in order to cut off the high Fourier components, since the original Green's function diverges at short distances (see Fig. 2). The inverse Fourier transform $\bar{G}(\rho)$ of $\tilde{g}(k)\, e^{-\alpha^2k^2}$ can be obtained analytically,

$$\begin{aligned} \bar{G}(\rho) = \frac{\sqrt{2\pi}}{4\rho}\, e^{\alpha^2\kappa^2}\Big[ & e^{\kappa\rho}\,\mathrm{erf}\Big(\alpha\kappa+\frac{\rho}{2\alpha}\Big)+ \\ & -e^{-\kappa\rho}\,\mathrm{erf}\Big(\alpha\kappa-\frac{\rho}{2\alpha}\Big)-2\sinh(\kappa\rho)\Big]. \end{aligned} \tag{11}$$

For each pair of tip and sample states we choose an $\alpha$ value large enough to ensure that the weight of the maximum representable $k$-components in the mesh are below a given tolerance, but sufficiently small not to perturb the shape of the tail at large distances, imposed by a maximum matching distance $\rho_m$ (Fig. 2). In all cases the coincidence is maintained for relative distances below 1 Bohr, so that the propagation of the values of wave functions through vacuum is not altered at all for separations in the tunnel regime.

### D. Matrix elements as convolutions

To obtain the final matrix elements, the propagated wave functions of Eq. (9) can be substituted into Eq. (2). After broadening the second surface of integration and by means of an additional application of Green's theorem, the following closed form can be written down

$$M_{ts}(\boldsymbol{R}) = -\frac{\hbar^2}{2m}\int[\boldsymbol{B}_t^*(\boldsymbol{r}'-\boldsymbol{R})\cdot\nabla\varphi_s(\boldsymbol{r}')-\varphi_s(\boldsymbol{r}')A_t^*(\boldsymbol{r}'-\boldsymbol{R})]d^3r' = \frac{\hbar^2}{2m}\int[\tilde{A}_t^*(\boldsymbol{k})-i\boldsymbol{k}\cdot\tilde{\boldsymbol{B}}_t^*(\boldsymbol{k})]\tilde{\bar{g}}_{ts}(k)[\tilde{A}_s(\boldsymbol{k})+i\boldsymbol{k}\cdot\tilde{\boldsymbol{B}}_s(\boldsymbol{k})]e^{i\boldsymbol{k}\cdot\boldsymbol{R}}d^3k, \tag{12}$$

where we have shifted all variables related with the tip up to position $\boldsymbol{R}$. The expression is very symmetric, which permits the factorization of tip and sample variables to simplify the coding and its evaluation. Hence, $M_{ts}$ can be evaluated for multiple locations of the tip in a sole computation by three-dimensional fast Fourier transforms, using values of $\boldsymbol{c}_j(\boldsymbol{r})$, $\varphi_j(\boldsymbol{r})$ and $\nabla\varphi_j(\boldsymbol{r})$, amassed in the points of an uniform grid within the regions of the broadened surfaces $\Sigma_j$.

### E. Energy integral

Lastly, the occupation factors and the energy delta functions in Eq. (1) can be rewritten as

$$\begin{aligned} &[f(\varepsilon_t)-f(\varepsilon_s)]\,\delta(\varepsilon_t-\varepsilon_s+eV) = \\ &\quad\int_{-\infty}^{+\infty}[f(\varepsilon)-f(\varepsilon+eV)]\,\delta(\varepsilon-\varepsilon_t)\,\delta(\varepsilon-\varepsilon_s+eV)\,d\varepsilon. \end{aligned} \tag{13}$$

Within this formulation the energy states of tip and sample are naturally decoupled in two functions which directly depend on the eigenvalues of the isolated systems. This is an important point, since in the electronic structure calculations the electrodes are represented by a finite number of states, so that the coupling with their respective bulks for the case of semi-infinite systems is not totally included. The effect of such contact will be essentially the shift and the broadening of the isolated energy levels [27], which is regularly treated through the inclusion of a self-energy matrix in the Hamiltonian of the finite system, playing the role of a large reservoir.

In the evaluation of Eq. (1), this is mimicked by smoothing the delta functions in Eq. (13), i.e., substituting them by Lorentzian functions,

$$\delta_{\eta_j}(\varepsilon-\varepsilon_j) = \frac{\eta_j}{\pi[(\varepsilon-\varepsilon_j)^2+\eta_j^2]}, \tag{14}$$

and introducing empirically fitted self-energy values $\eta_j$. Although in principle they should depend on various parameters of the calculations, e.g., system size, $k$-sampling, or the coupling of each state to the bulk, we have chosen the simplistic criterion of using the lowest constant self-energy for each electrode such that the simulated $I-V$ curves do not present negative differential conductances. In practice, since the tip will be represented by fewer states, and therefore more sensitive to the broadening, the criterion is achieved by setting a reasonable value for the sample (of about tens of meV, in view of its energy distribution around the Fermi level), and adjusting it for the probe.

The computation of the integral in Eq. (13) can be also implemented using convolutions and fast Fourier transforms, although a fine grid in real space is necessary for a good sampling of the Lorentzian functions. Alternatively, the integral can be performed applying the Sommerfeld expansion [28], which up to second order in the electronic temperature reads as

$$\int_{-\infty}^{+\infty} [f(\varepsilon) - f(\varepsilon + eV)]\, \delta_{\eta_t}(\varepsilon - \varepsilon_t)\, \delta_{\eta_s}(\varepsilon - \varepsilon_s + eV)\, d\varepsilon \approx \\ \approx K_0(eV) + K_2(eV, kT_t, kT_s), \tag{15}$$

where $K_0$ and $K_2$ are the following functions,

$$K_0 = D^{-1} \Big\{ \eta_s[\eta_t^2 - \eta_s^2 - (\varepsilon_t - \varepsilon_s + eV)^2] \arctan\left(\frac{\varepsilon_t}{\eta_t}\right) + \\ + \eta_t[\eta_t^2 - \eta_s^2 + (\varepsilon_t - \varepsilon_s + eV)^2] \arctan\left(\frac{\varepsilon_s}{\eta_s}\right) + \\ + \eta_s[\eta_s^2 - \eta_t^2 + (\varepsilon_t - \varepsilon_s + eV)^2] \arctan\left(\frac{\varepsilon_t + eV}{\eta_t}\right) + \\ + \eta_t[\eta_s^2 - \eta_t^2 - (\varepsilon_t - \varepsilon_s + eV)^2] \arctan\left(\frac{\varepsilon_s - eV}{\eta_s}\right) + \\ + \eta_t\eta_s(\varepsilon_t - \varepsilon_s + eV)\left[\log\left(\frac{\eta_t^2 + (\varepsilon_t + eV)^2}{\eta_t^2 + \varepsilon_t^2}\right) + \\ + \log\left(\frac{\eta_s^2 + (\varepsilon_s - eV)^2}{\eta_s^2 + \varepsilon_s^2}\right)\right]\Big\},$$

$$D = \pi^2[(\eta_t + \eta_s)^2 + (\varepsilon_t - \varepsilon_s + eV)^2] \times \\ \times [(\eta_t - \eta_s)^2 + (\varepsilon_t - \varepsilon_s + eV)^2], \tag{16}$$

$$K_2 = \frac{\pi^3}{3} \Big\{(kT_t)^2 \delta_{\eta_t}(\varepsilon_t)\delta_{\eta_s}(\varepsilon_s - eV)\, [\delta_{\varepsilon_t}(\eta_t) + \delta_{\varepsilon_s - eV}(\eta_s)] + \\ - (kT_s)^2 \delta_{\eta_t}(\varepsilon_t + eV)\delta_{\eta_s}(\varepsilon_s)\, [\delta_{\varepsilon_t + eV}(\eta_t) + \delta_{\varepsilon_s}(\eta_s)]\Big\}. \tag{17}$$

### F. Practical simulation parameters

The scheme presented in the previous sections involves the use of some parameters that must take values within an appropriate range. For instance, the mathematical surfaces of integration, $\Sigma$, defined by Eqs. (6) and (7), are controlled via $\rho_0$ and $\Delta S$. In practice, we have defined $\rho_0$ according to the electron radius, $r_s = (3/4\pi\rho_0)^{1/3}$, and we have chosen values $r_s \sim 8 - 10$ Bohr, following the condition argued in section II B. In the ensuing section III, we have checked the dependence of the simulated images against variations in $r_s$, finding that in these cases there is almost no dependence. Nonetheless, there might exist situations (e.g., adsorbates that modify the local work function), where the reference density should be chosen within a tighter window. The value of the parameter $\Delta S$ is estimated assuming that the electron density in the $u$-direction behaves as $\rho(u) \sim \exp(-2\kappa u)$. If we make $\Delta S \gtrsim 2\pi\sqrt{\phi/E_{\text{cut}}}$, there will be at least one point of the grid (whose spacing is defined by the energy cutoff, $E_{\text{cut}}$, of the representable plane waves), sampled in that local perpendicular direction. Using typical values of the work function $\sim 5$ eV, we see nearly no dependence of the results with values of $\Delta S$ up to 50% higher than the estimation. We have checked that variations in $\phi$ cause changes in the absolute value of the total tunneling current, but the apparent heights and corrugations in the surface remain practically the same.

The $\eta_j$ parameters in the present implementation of the energy broadening can be more tricky to adjust. Generally, the energy distribution of the sample represented by a slab will resemble much more that of a semi-infinite system than that of a tip described using only tens of atoms. The criterion established in section II E leads to self-energies of about 1 − 1.5 eV for the tips, in order to smooth appropriately the HOMO-LUMO gap present in the isolated cluster.

Finally, since this method is mainly designed for simulations using local basis sets, it is important to assess the importance on the basis functions. In our calculations we have systematically employed a double-$\zeta$ plus polarization (DZP) basis set and performed some checks with a smaller single-$\zeta$ (SZ) basis set. We generally find that the number of basis orbitals and their spatial extension is not critical, because the wave functions need to be accurate only at the surfaces $\Sigma$, which are relatively close to the atoms.

## III. SIMULATIONS

In the following we present simulations and comparisons with experiments for the Si(111)-(7×7) and Ge(111)-$c$(2×8) surfaces carried out in ultrahigh vacuum with tungsten tips. Details about sample preparation and experimental conditions are described for each system in Refs. [29] and [30] respectively.

The wave functions were obtained within density-functional theory [31] (in the local density approximation of Perdew and Zunger [32]) in all cases, both for samples and tips. We use the numerical atomic-orbital method in the

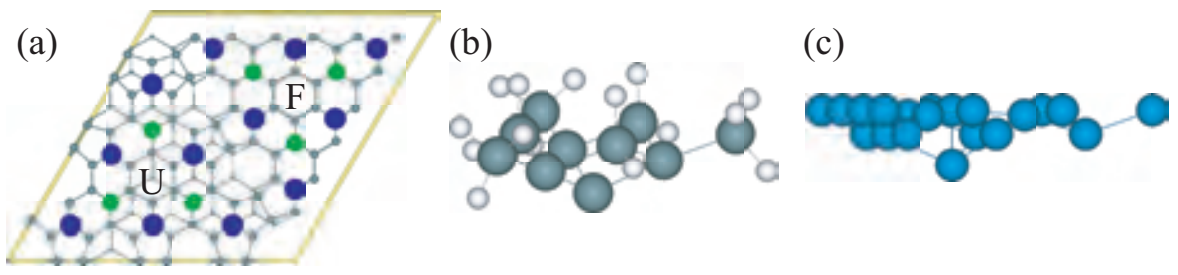

FIG. 3: Ball-and-stick models. (a) Faulted (F) and unfaulted (U) half-cells in the Si(111)-(7×7) unit cell (yellow). Ad-atoms are colored in blue, rest-atoms in green. (b) Tip formed by 10 silicon atoms saturated in the base. (c) W(111) bcc pyramid tip.

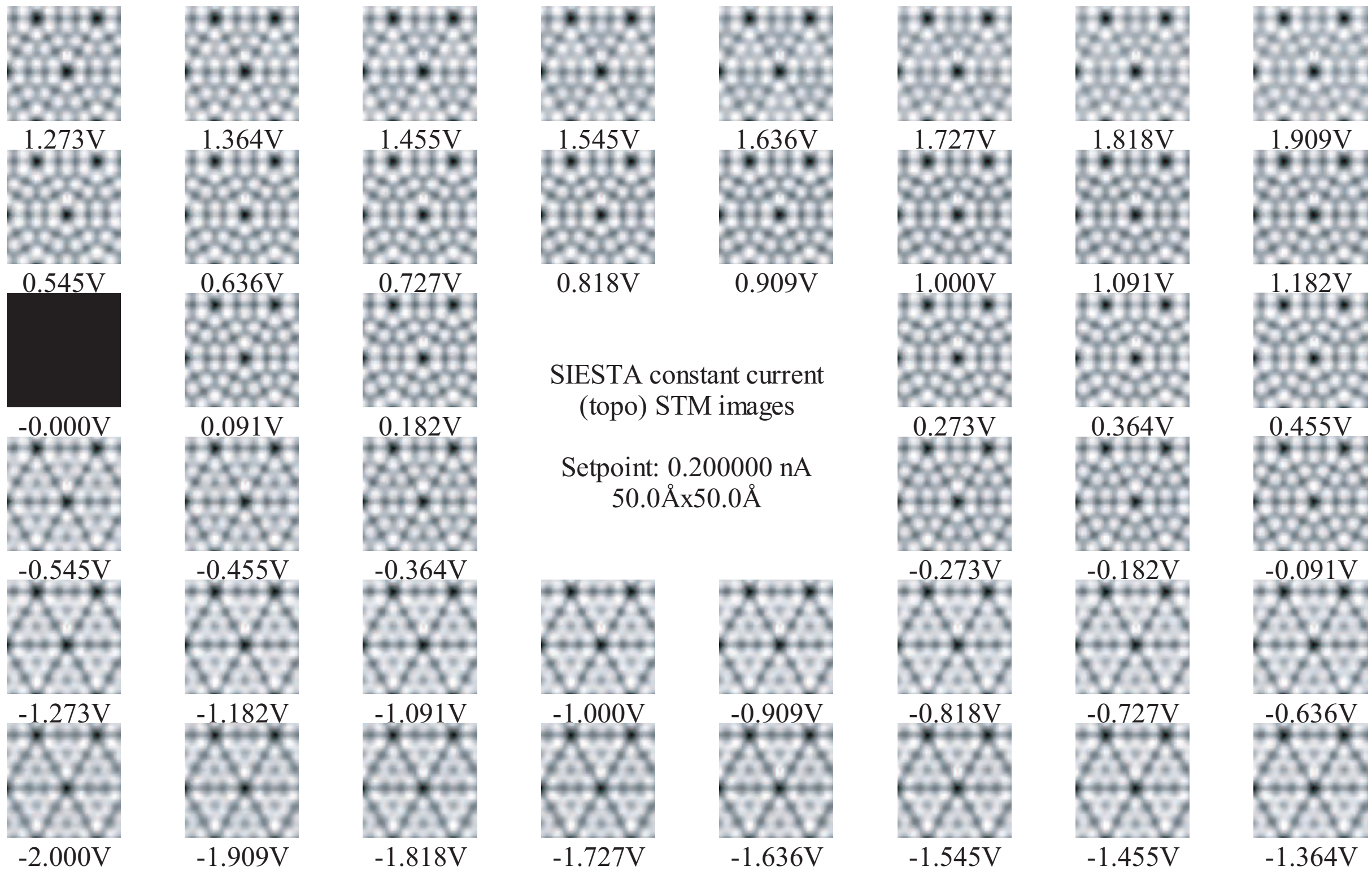

FIG. 4: Simulated constant-current topographies for the Si(111)-(7×7) surface at multiple voltages and using the Si tip (Fig. 3), as seen in the output window of the visualization program WSxM [36]. The set-point is fixed at 0.2 nA.

Siesta code implementation [33, 34], where core electrons are replaced by norm-conserving pseudo-potentials [35], whereas valence electrons are described using a DZP basis set. A real-space grid with a plane-wave cutoff of 100 Ry was set in all systems to compute the Hartree and the exchange-correlation contributions to the self-consistent potential and the Hamiltonian matrix, and to project the final wave functions $\varphi_s$ and $\varphi_t$. Only the Γ-point in the reciprocal space was used due to the large unit cell of the surfaces. The systems were relaxed independently until the maximum residual force was below 0.04 eV/Å. The values of the tunnel current $I(\boldsymbol{R}, V)$, calculated for each tip as explained previously, were dumped on files which were then read by the experimental data-acquisition program WSxM [36] and processed in exactly the same way as the experimental data.

### A. Si(111)-(7×7) surface

Structure relaxations from first-principles methods of the Si(111)-(7×7) surface reconstruction have been carried out in the past [37–41]. The major challenge resides in dealing with a large unit cell which contains hundreds of atoms (Fig. 3). Even so, this system presents a good experimental reproducibility and a rich variety of topographic and spectroscopic characteristics [5], making it an ideal benchmark for STM/STS simulations. In this work, the calculations were performed using a repeated slab geometry with four layers of silicon (the lowest of them saturated with hydrogen atoms), i.e., a total of 249 atoms. The simulation cell comprises a vacuum region in the perpendicular direction large enough both to ensure no interaction between periodical images of the sample and to include the volume of the tips used. We consider two tips: the first was made of ten silicon atoms in the configuration of (111) planes, proposed by Pérez *et al*. [42], in which all the dangling-bonds except that of the apex are saturated with hydrogen atoms; the second tip was a tungsten bcc pyramid pointing in the (111) direction, with 20 atoms (Fig. 3).

Figure 4 shows the simulation of STM images in the constant-current mode using the Si tip and for sample bias voltages in a range of ±2 V. Despite the very large unit cell of this surface, the typical CPU time for the simulations (without including the geometry relaxations) took only about 1 hour in a single processor PC. It includes the computation of currents for tip positions at all mesh points in the cell and for all voltages shown in the picture. Furthermore, the check with a SZ basis set, which is accomplished very quickly, generated nearly identical images, showing that if the quality of wave functions is not critical, reliable results could be achieved almost in real time. In Fig. 4, at first sight some of the most characteristic features of this surface can be observed both in empty and occupied STM images. The picture shows clear evidences in the difference between the apparent heights of the faulted and unfaulted half-unit cells for negative sample voltages, which cannot be perceived so patently in the empty state images. Other interesting aspect is the observation of the rest-atoms in the occupied state images (for bias voltages lower than $\sim -0.5$ V) that cannot be resolved at positive sample voltages. The simulation employing a W tip (not shown here) follows the same tendency except for presenting a less pronounced contrast between both half-cells, an effect that has been also appreciated by experimentalists. In fact, a common practice in order to enhance the experimental resolution is to intentionally perform slight tip-sample contacts before scanning, which justifies the widespread assumption that the effective STM tip (originally W wire in these measurements) can be terminated on a Si cluster.

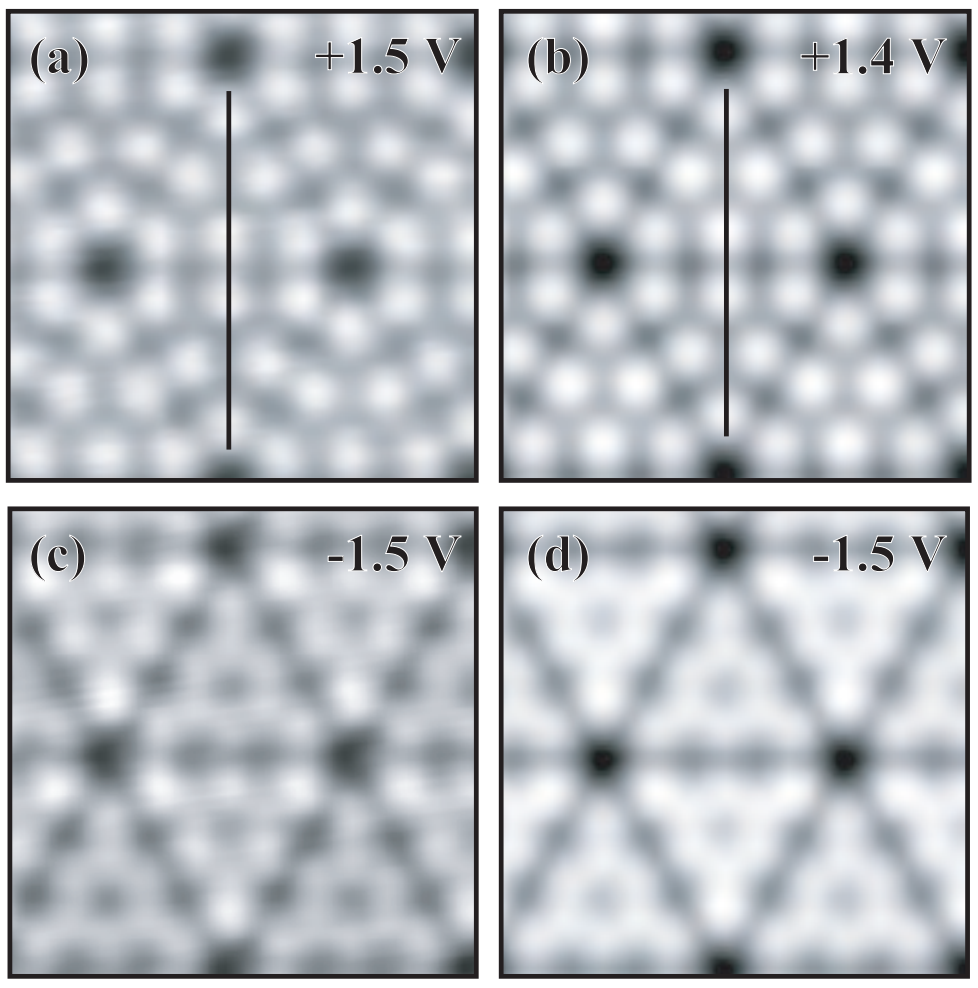

FIG. 5: Experimental constant-current STM topographic data (left panels) and simulations with a Si tip (right panels) of the Si(111)-(7×7) surface at 0.2 nA. (a),(b) Empty state and (c),(d) occupied state images using the same gray scale. Reproduced with permission from Ref. [29].

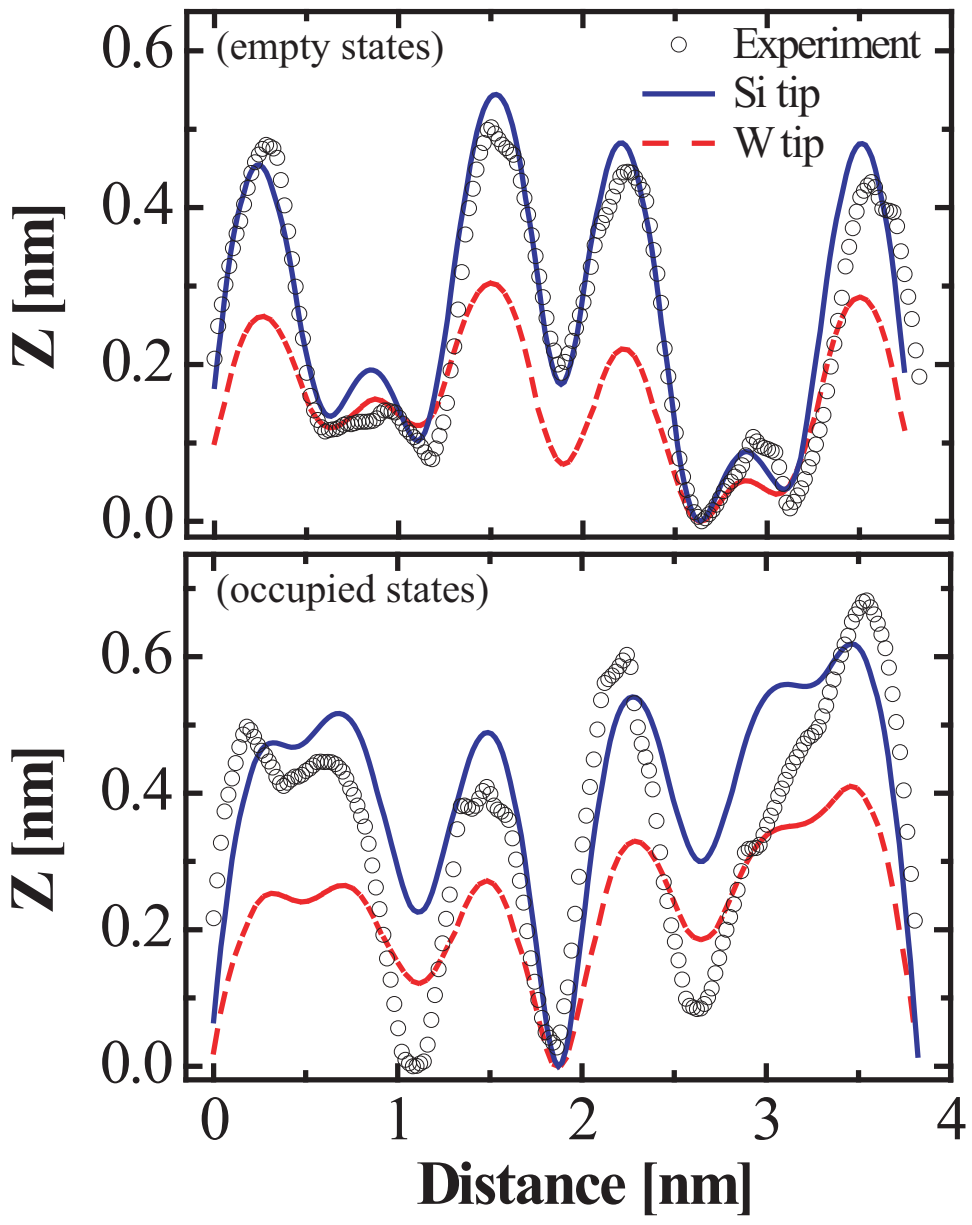

FIG. 6: Profile comparison between experiment and theory for positive (up) and negative (down) sample voltages, along the solid lines in Fig. 5. Curves have been shifted to make their minima coincide.

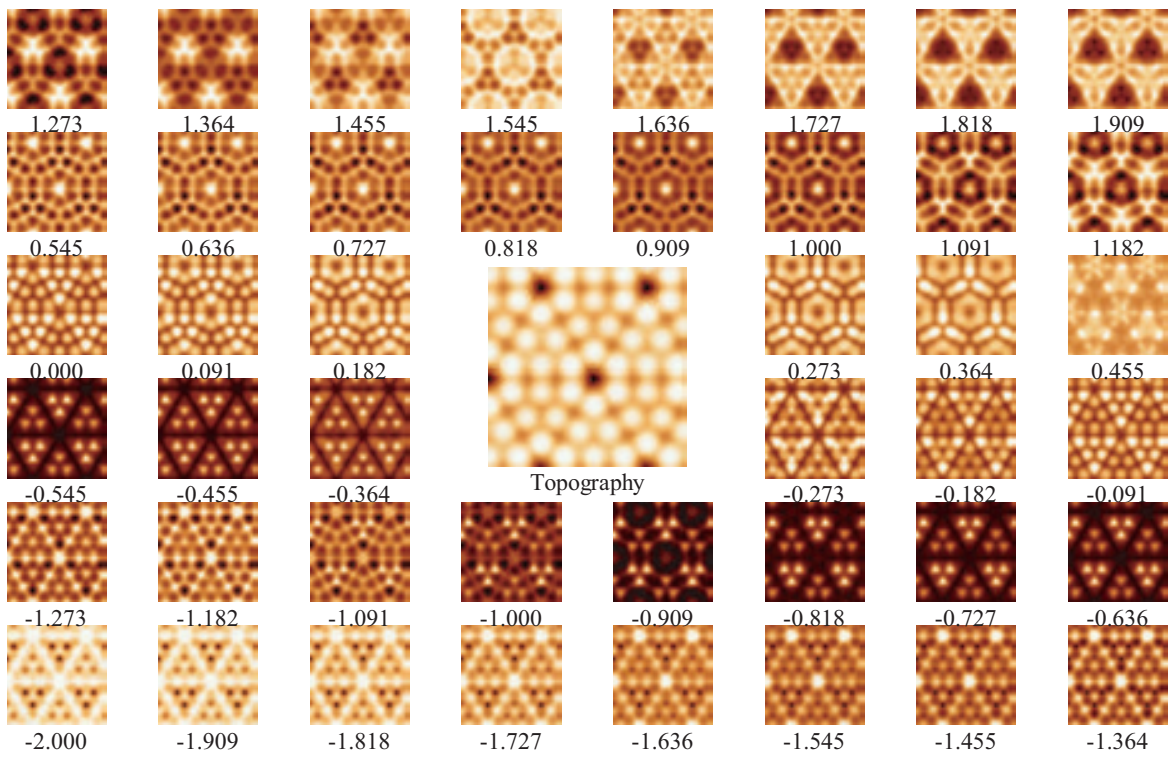

FIG. 7: Theoretical data for CITS in the case of a silicon tip. The central picture shows the topographic image at the set-point of 2 nA and 1.73 V. The remaining images represent $\partial I/\partial V$ as a function of $V$, in Volts, at the tip-sample distance determined by the set-point.

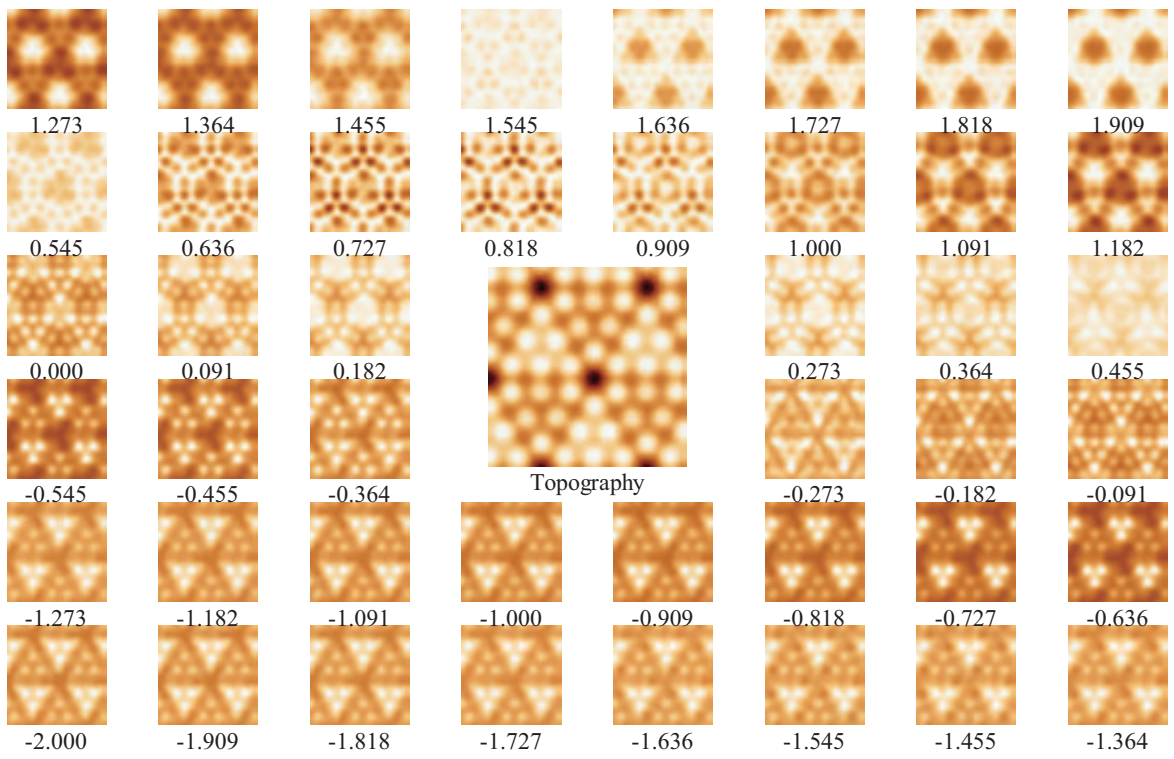

FIG. 8: Same as Fig. 7 for the case of a W(111) tip.

A comparison with experiments is presented in Fig. 5. The experimental data correspond to typical empty (+1.5 V) and occupied (−1.5 V) STM images, measured under the experimental conditions described in Ref. [29], and usually obtained after touching lightly the surface with the tip. The simulated images shown in Fig. 5 are those of Fig. 4 with similar tunnel parameters (bias voltage and current) as the experimental ones [43]. The calculated tip-sample distance in the simulations is about 7 Å. There is a very good overall agreement between experimental and theoretical topographic images. Only some small differences can be noticed in occupied state images, where the apparent height increment between faulted and unfaulted half-cells is larger in the experimental data than in the simulation. However, the difference could be interpreted as a structural effect caused by an imperfect surface relaxation, due to the limitation in using only four layers of silicon to mimic the bulk, as pointed out by Ref. [39]. This also suggests us to compare the empty state image with simulated data close to the experimental bias voltage [43]. Differences in the topography between simulations and experiments are more clearly observed in profile lines. The quantitative comparison for a path crossing both halves in the unit cell is shown in Fig. 6. From this data it is concluded that the agreement with experiments (open circles) is considerably better for the Si tip (solid line), specially at positive sample voltages, and worse for the W tip (dashed line). In the former, the corrugation is appreciably higher, in agreement with the experimental observations after shaping the tip by slightly contacting the sample, and strongly suggesting that the chemical nature of the tip could have been modified during operation with highest resolution.

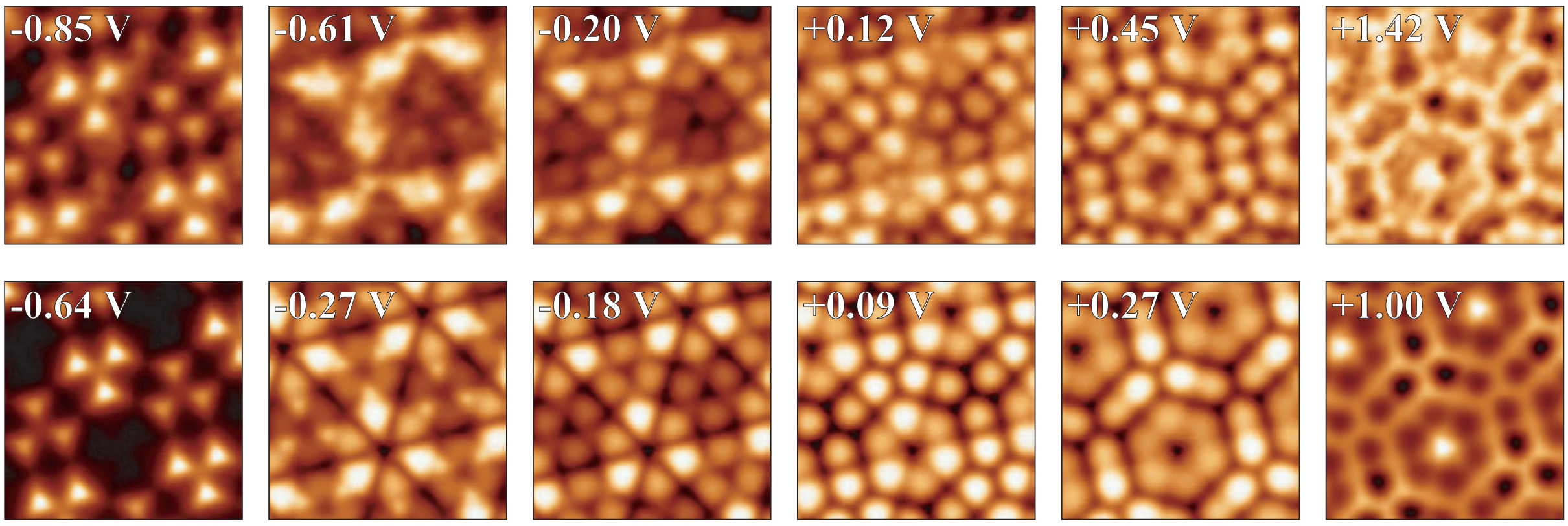

FIG. 9: CITS comparison between experiments (top pictures) and simulations using a silicon tip (bottom pictures) extracted from Fig. 7. Reproduced with permission from Ref. [29].

A further discrimination between the two proposed tips can be expected from a spectroscopic analysis of the sample. Experimental data were taken in the current-imaging-tunneling spectroscopy (CITS) mode [5], where the topography $z(x, y)$ is scanned in the constant-current mode and the feedback is eventually looped off in every point $\mathbf{(}x, y, z(x, y)\mathbf{)}$ to acquire a current vs voltage $(I-V)$ curve. Thus, $I(x, y, V)$ is recorded for a fixed control current and voltage at many different bias voltages and coordinates, and CITS maps of $\partial I/\partial V$ obtained from direct numerical differentiation [36]. Figures 7 and 8 show the CITS maps using the silicon and the tungsten tip respectively, where many different patterns are observed as the voltage changes. In the case of a Si tip, the images appear with more contrast than for the W tip and a richer family of patterns is observed for negative bias voltages. This region of energies is specially sensitive to the DOS of the tip, since the predominant current at this voltages occur from electrons which tunnel, through a lower barrier, from surface electron states at the Fermi level to empty states in the tip.

Motivated by the previous agreements in the topographies, in Fig. 9 we compare with experiments some representative pictures from Fig. 7, chosen in various ranges of voltages for which the same pattern is observed in the simulations. Although they appear at different bias voltages, all the very different experimental patterns are closely reproduced. The agreement is much worse using a tungsten tip. Therefore, it is safe to conclude that, for these measurements, the shape and the chemical composition of the tip was modified. As a result of the modification, it was very probably terminated with Si, with a configuration similar to that of the tip employed in the simulations.

### B. Ge(111)-$c(2\times8)$ surface

*Ab initio* calculations of this surface were previously reported [40, 41, 44]. This reconstruction has also attracted attention regarding the diffusion of surface vacancies [30, 45]. In the present work, the calculations were performed following the same scheme as that used in section III A. The surface was mimicked by a slab with six layers of Ge and with an additional layer of hydrogens in the bottom. The unit cell has 74 surface atoms and a volume large enough to fit the tip into the vacuum space left in the nonperiodic direction of the simulation cell. Based on the experience with the Si(111)-(7×7) surface, the tip employed in this calculations is formed by a cluster of Ge atoms, in the same configuration of the Si tip used previously, represented in Fig. 3(b).

Figure 10 exhibits the STM simulation of topographic images for empty and occupied state images. In this case, there apparently exist only two qualitatively different images. For positive sample voltages above $\sim$ 0.5 V, ad-atoms are predominantly seen, and only some small traces from the rest-atoms can be observed. In the rest of the images, the most superficial atoms are seen intermingled, which generates a family of slightly different patterns, as the relative contributions from the ad-atoms and rest-atoms change with the bias voltage.

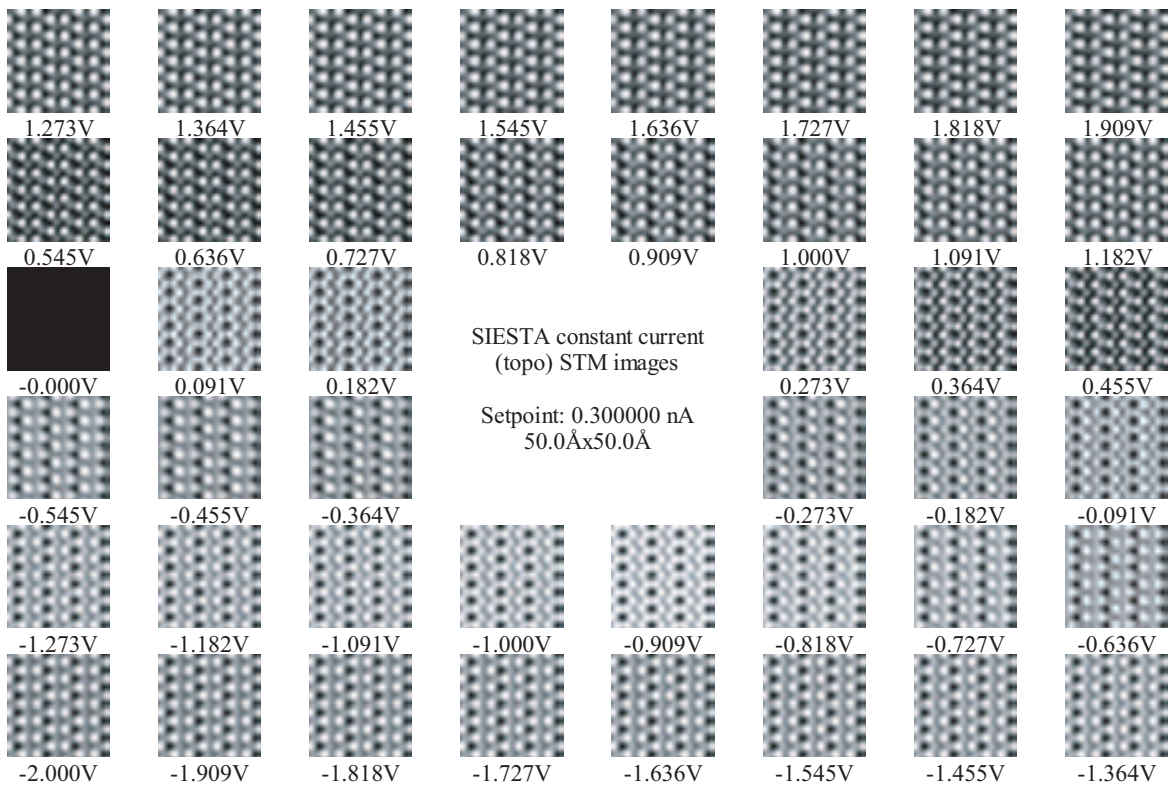

FIG. 10: Simulations for the Ge(111)-$c(2\times8)$ surface using a Ge tip with a similar structure to that in Fig. 3(b). The picture shows a snapshot of the visualization program [36], presenting topographies in the constant-current mode at 0.3 nA and for bias voltages in the range of ±2V.

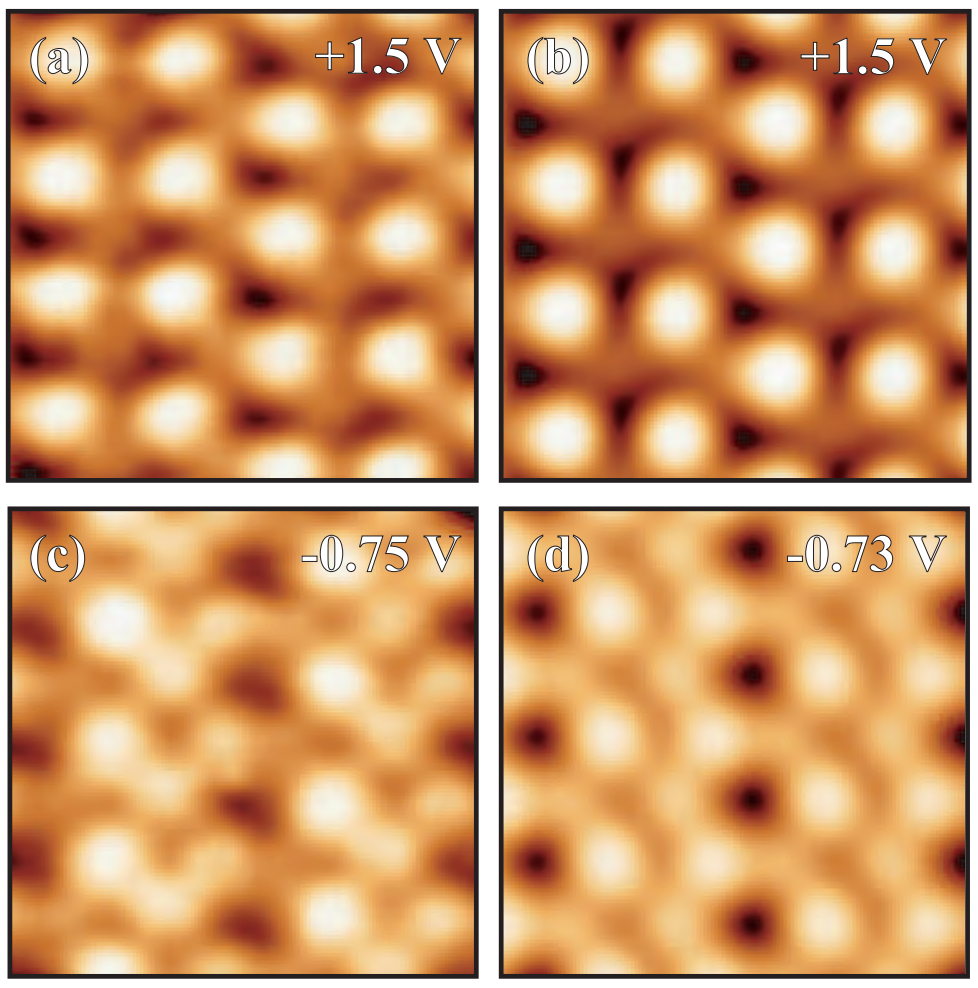

FIG. 11: (a),(b) Empty state and (c),(d) occupied state images at 0.3 nA. The left panels show measurements, while the right panels represent simulations using a Ge tip (see the text).

In Fig. 11, experimental images of the topography (at the constant current of 0.3 nA) are compared with those of the theory extracted from Fig. 10. As observed, the agreement is fairly good for both bias voltages. At the negative sample voltage, the likeness between simulations and experimental data can be considered more clearly, since the same combination of ad-atoms and rest-atoms has been always reproduced in different measurements. This is not the case for the empty state image, however, where not always the same picture was found in the experiments, and sometimes recordings without any sign of rest-atoms have been also appreciated. This fact emphasizes the difficulty to determine conclusively the shape of the tip in all situations, although strongly suggests that the chemical composition of the tip could be also contaminated with germanium.

## IV. CONCLUSIONS

In summary, we have developed an accurate method for fast simulation of STM images at realistic experimental tip-sample distances from first-principles calculations of the surface and tip at the same footing. The method is based in the propagation of wave functions across vacuum and allows the efficient use of basis sets of atomic orbitals to describe each system independently. All the tunneling currents for different bias voltages and tip positions are obtained in a single convolution and very quickly using fast Fourier transforms. It permits to achieve reliable results using tips with multiple configurations or compositions, that can be compared directly with experiments. In our findings, comparisons with two proposed tips show that the original experimental W tip was contaminated during the scanning. In the analysis of the Si(111)-(7×7) surface, we ascertained that the effective STM tip should resemble a cluster formed by silicon atoms. A similar tip, made up of Ge atoms, suggests the same situation for some measurements of the Ge(111)-*c*(2×8) reconstruction.

### Acknowledgments

We are indebted to I. Brihuega and J. M. Gómez-Rodríguez for fruitful discussions, STM and STS measurements and for their help in the visualization and comparison of the theoretical data with experiments. This work has been supported by Grant No. BFM2003-03372 from the Spain's Ministry of Science, and by a FPI studentship from the Autonomous Community of Madrid.